\documentclass[preprint,epj]{svjour}
\usepackage{graphicx}
\begin{document}
\title{Isotope shifts of 6s5d $^3$D-states in neutral
barium}

\author{U. Dammalapati\thanks{Present address: University of Strathclyde, Glasgow, UK},
S. De, K. Jungmann, and L. Willmann}

%\

\institute{Kernfysisch Versneller Instituut, University of
Groningen, Zernikelaan 25, 9747 AA Groningen, The Netherlands}

\date{\today}

\abstract{ Laser spectroscopy of the low lying $^1$P and $^3$D
states in atomic barium has been performed. This work contributes
substantially to the development of an effective laser cooling and
trapping for heavy alkaline earth elements and aims in particular
for a better understanding of the atomic wave function of these
systems. Isotope shifts and hyperfine structures are ideal probes
for the wave functions at the position of the nucleus. This is
essential input for a theoretical evaluation of the sensitivity to
fundamental symmetry breaking properties like permanent electric
dipole moments. We report the first isotope shift measurements of
the $^3$D$_{1,2}$-$^1$P$_1$ transitions. A deviation of the King
plot from its expected behavior has been observed. Further we have
optically resolved the hyperfine structure of the $^3$D$_{1,2}$
states.}


\authorrunning{U. Dammalapati et al.}
\titlerunning{Isotope shifts of 6s5d $^3$D-states in Ba}

\maketitle

\section{Introduction}

Recently heavy alkaline earth atoms have attracted attention due to
high enhancement factors for possible permanent electric dipole
moment (edm's), both of electrons and nuclei in several isotopes
which have nuclear spin. Of particular interest are radium (Ra)
isotopes which exhibit the largest known enhancement factors for an
electron edm and for nucleon edm in nuclei. These arise from close
lying states of opposite parity \cite{FLAMBAUM1999,FLAMBAUM2000} in
the atomic shell ($^3$P and $^3$D) or in the nucleus where they are
associated with octopole deformations in some nuclei near the region
of the valley of stability \cite{engel,flambaumoct}. In order to
conduct such research a good understanding of the atomic structure,
i.e. wave functions, transition probabilities, lifetimes, and
hyperfine structure splitting, is needed. At this time atomic
structure calculations with independent numerical approaches are
underway \cite{privcomm}.

Sensitive experiments with rare or radioactive isotopes such as Ra
require samples of laser cooled and trapped atoms. Experiments are
prepared at Argonne National Laboratory, Il, USA \cite{guest2007}
and at the TRI$\mu$P facility at the Kernfysisch Versneller
Instituut, The Netherlands \cite{jungmann1,Jungmann2002}. The
development of an effective method for laser cooling and trapping is
indispensable. Elaborate laser cooling schemes need to be explored.
They involve several lasers at the same time. Barium (Ba), a
chemical homologue to Ra, has a similar level structure and is well
suited for the development of the technical method due to the
copious availability of a large number of stable isotopes.

\begin{figure}[tb]
\center
\includegraphics[width = 75mm, angle = 0]{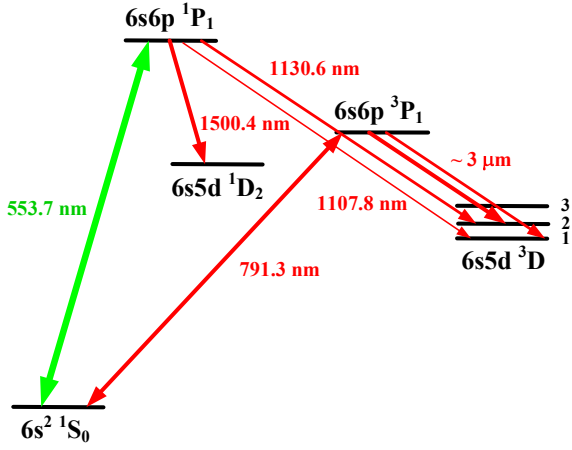}\\
\caption{Level diagram of the Ba atom. The wavelengths for the
transitions used in laser spectroscopy of the metastable D-states
are given.} \label{figure1}
\end{figure}

In atomic Ba, the 6s$^{2}$~$^{1}$S$_{0}$~-~6s6p~$^{1}$P$_{1}$
transition (Fig.~\ref{figure1}) is the strongest transition. It is
the only option for laser cooling and trapping in the ground state.
The upper state branches into the 6s5d $^{1}$D$_{2}$, $^{3}$D$_{2}$
and $^{3}$D$_{1}$ states. This transition has been exploited in the
past successfully to determine nuclear magnetic moments and isotope
shifts \cite{WIJNGAARDEN1995}. The level scheme of Ba (Fig.
\ref{figure1}) is characterized by low lying $^1$D and $^3$D-states.
This causes a rather large decay of the $^1$P$_1$ state into all of
these states and puts a strong constraint on the optical cooling
cycles. The known transition probabilities (Tab.
\ref{refwavelength}) yields a branching ratio of 1:330(30) for Ba.
This value is 1:350(50) for Ra. In contrast, the branching is
1:40000 for the next lighter alkaline earth element strontium, and
therefore allows for laser cooling and trapping with a single laser
frequency \cite{katori}.

The similarity in the atomic structure of Ba and Ra allows to work
with Ba to establish the experimental techniques and the theoretical
framework relevant for searches of edm's in Ra. This paper focusses
on the transitions leading to losses from the $^1$S$_0$-$^1$P$_1$
cooling cycle by branching to metastable D-states. We optically
resolved the hyperfine structure of the $^3$D-states and measured
several isotope shifts. In particular the isotope shifts reveal
important correlation effects of the electrons, which are relevant
for evaluating enhancement factors for edm's. The developed
techniques will allow to perform such measurements with the much
less abundant Ra isotopes.

\begin{table}[tb]
\begin{center}
\begin{tabular}{|c|c|c|r|l|}
\hline
Upper  &  Lower & Label &$\lambda$ & A$_{ik}$ \\
Level (i) & Level (k) & & [nm] & [10$^{8}$ s$^{-1}$] \\
\hline
$^{1}P_{1}$ &   $^{1}S_{0}$ & $\lambda_1$ & 553.7 & 1.19(1)$^{a}$\\
 & $^{1}D_{2}$  & $\lambda_{ir1}$ & 1500.4 & 0.0025(2)$^{a}$\\
    &  $^{3}D_{2}$  & $\lambda_{ir2}$ & 1130.6 & 0.0011(2)$^{a}$\\
 &  $^{3}D_{1}$  & $\lambda_{ir3}$ & 1107.8 & 0.000031(5)$^{a}$ \\
\hline
$^{3}P_{1}$ & $^{1}S_{0}$ & $\lambda_2$ & 791.3 & 0.0030(3) \\
& $^{3}D_{2}$ & & 2923.0 & 0.0032(3)$^{b}$\\
& $^{3}D_{1}$ & & 2775.7 & 0.0012(1) $^{b}$\\
\hline
\end{tabular}
\end{center}
\label{refwavelength} \caption{The energy levels, vacuum wavelengths
$\lambda$ and transition rates of the Ba atom relevant for the
experiments. Experimental transition rates A$_{ik}$ are taken from:
`a' \cite{NIGGLI1987,BIZZARI1990}, `b' \cite{BRUST1995}.}
\label{bawavelengths}
\end{table}

\begin{table}
\begin{tabular}{|c|c|c|c|c|}
\hline
State & $^{1}P_{1}$ & $^{1}D_{2}$ & $^{3}D_{2}$ & $^{3}P_{1}$ \\
\hline $\tau$ & 8.2(2)ns\cite{LURIO1964} &
 0.25s\cite{MIGDALEK1990} &
 60s\cite{MIGDALEK1990} &
 1345(14)ns\cite{SCIELZO2006}\\
 \hline
\end{tabular}
\caption{ Lifetimes of the $^1$P$_1$, $^1$D$_2$, $^3$D$_2$ and
$^3$P$_1$ states. The reference \cite{LURIO1964,SCIELZO2006} quote
experimental values and \cite{MIGDALEK1990} quote calculated
values.}
\end{table}

\section{Experimental approach}

An effusive Ba atomic beam is produced from Ba metal at about 800 K
in a resistively heated oven. The abundances of naturally occurring
Ba isotopes are $^{138}$Ba (71.7\%), $^{137}$Ba (11.3\%), $^{136}$Ba
(7.8\%), $^{135}$Ba (6.6\%), $^{134}$Ba (2.4\%), $^{132}$Ba (0.1\%)
and $^{130}$Ba (0.1\%). The ground state nuclear spin of the even
isotopes is zero and for the odd isotopes it is 3/2 $\hbar$.

\begin{figure}[tb]
\center
\includegraphics[width = 55 mm, angle = 270]{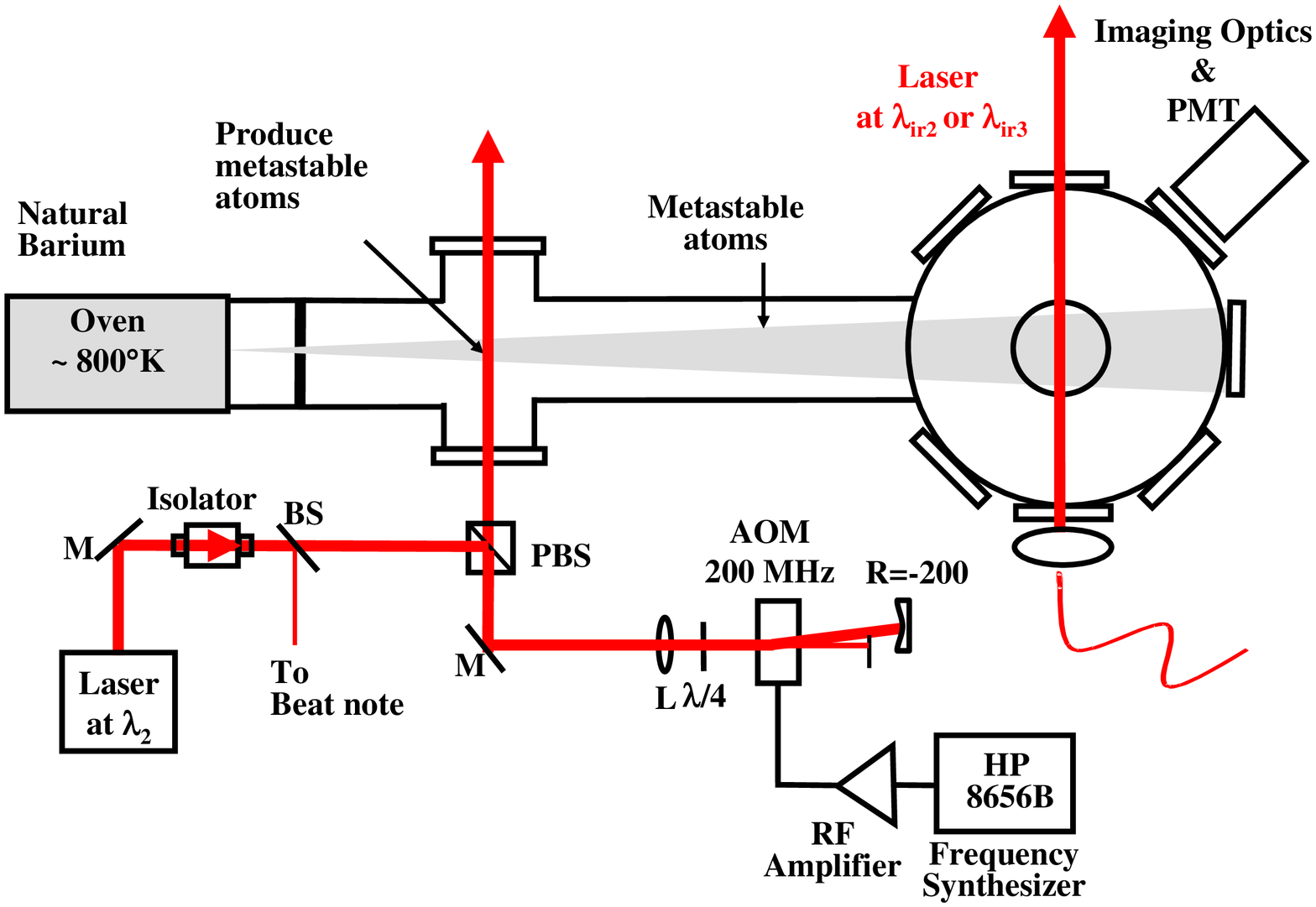}\\
\caption{Setup for isotope shift measurements using a beam of
metastable Ba atoms. In the first interaction zone, atoms are
optically pumped into metastable D-states with  laser light at
$\lambda_2$=791.3~nm. A second interaction zone allows to probe the
metastable atoms with the infrared lasers. The transitions are
detected by fluorescence at $\lambda_1$ from the
$^{1}$P$_{1}$-$^1$S$_0$ transition with a photo multiplier.}
\label{figure2}
\end{figure}

The experimental apparatus (Fig. \ref{figure2}) is designed to
fulfill two requirements: First, the production of an atomic beam of
metastable D-state atoms by optical pumping. Second, infrared laser
spectroscopy of the metastable states. The setup provides two
separate interaction zones of the atomic beam with laser beams. We
observe the transitions originating from the metastable D-states by
fluorescence at the wavelength $\lambda_1$. In this paper we focus
on the $^3$D states.

\subsection{Isotope selective metastable production}

Metastable atoms in the D-states are produced by optical pumping
with a laser beam orthogonal to the atomic velocity direction.
Isotope selective production can be accomplished either by driving
the 6s$^{2}$ $^{1}$S$_{0}$~-~6s6p $^{1}$P$_{1}$ transition at
wavelength $\lambda_1$~=~553.7~nm followed by spontaneous decay or
by the 6s$^{2}$ $^{1}$S$_{0}$~-~6s6p $^{3}$P$_{1}$ intercombination
line at wavelength $\lambda_2$~=~791.3~nm. The latter possibility
allows to populate the $^3$D-states exclusively. The isotope shifts
(Tab. \ref{isoshiftpump}) of these transitions permit isotope
selective optical pumping into the metastable state. We use the
second option because of the large branching of the $^3$P$_1$ state
to the $^3$D-states (Tab. \ref{bawavelengths}). In addition, the
narrow linewidth of the intercombination line at wavelength
$\lambda_2$ allows for a superior isotope selection. Among other
transitions, the $^{3}$D$_{2}$~-~$^{1}$P$_{1}$ transition at
wavelength $\lambda_{ir2}$=1130.6~nm and the
$^{3}$D$_{1}$~-~$^{1}$P$_{1}$ transition at wavelength
$\lambda_{ir3}$=1107.8~nm have been observed in fluorescence in
Fourier transform spectroscopy using a hollow-cathode discharge lamp
\cite{NIGGLI1987,BIZZARI1990}.

\begin{table}[tb] \center
\begin{tabular}{|c|c|c|}
  \hline
  Isotope pair & Isotope shift [MHz]& Isotope shift [MHz] \\
    & at $\lambda_1$ & at $\lambda_2$ \\
  \hline
  138-137 & -215.15(16) &-183.4(1.0) \\
  138-136 & -128.02(39) &-109.2(1.0) \\
  138-135 & -259.29(17) &-219.9(1.0) \\
  138-134 & -143.0(5) &-122.3(2.5) \\
%  138-130 & x &-174.7(1.2) \\
  \hline
\end{tabular}
\caption{Isotope shifts of the transitions at wavelength $\lambda_1$
\cite{WIJNGAARDEN1995} and $\lambda_2$ \cite{GRUNDEVIK1983} in
Ba.}\label{isoshiftpump}
\end{table}

We use two home built grating stabilized diode lasers at wavelength
$\lambda_2$. The Hitachi 7851G laser diodes deliver 15mW each. One
reference laser is stabilized against the 'a1' line of the
P(52)(0-15) transition in molecular iodine $^{127}$I$_{2}$ by
frequency modulation saturated absorption spectroscopy
(Fig.\ref{iodinespec}). The second diode laser is frequency offset
locked to the iodine stabilized laser. The offset frequency is
measured by a beat note on a fast photodiode. The beat frequency for
the different isotopes ranges from -3~GHz to 2~GHz because of the
large hyperfine structure of the odd isotopes. The light of this
second laser frequency shifted by double passing it through an
200~MHz acousto optical modulator (AOM) before the optical pumping
region. There we have typically 3-5~mW of laser power available. The
radio-frequency $\nu_{AOM}$ to the AOM is provided by a HP~8656B
frequency synthesizer. Rapid switching between different isotopes is
achieved by changing  $\nu_{AOM}$. The typical switching times are
limited by the synthesizer to around 50~ms. This procedure reduces
the systematic uncertainty of isotope shift measurements due to slow
drifts.

The stabilized diode laser system at wavelength $\lambda_2$ allow to
populate the $^3$D-states selectively. Around 60$\%$ of the Ba atoms
decay to the $^{3}$D$_{2}$ and $^{3}$D$_{1}$ states and 40$\%$ decay
back to the ground state. Cycling the $^{1}$S$_{0}$~-~$^{3}$P$_{1}$
transition populates the $^3$D-states with a ratio of 70$\%$ in the
$^{3}$D$_{2}$ and 30$\%$ in the $^{3}$D$_{1}$ states
(Tab.~\ref{bawavelengths}). At the saturation intensity of 65
$\mu$W/cm$^2$ an interaction length of about 30~mm is sufficient to
transfer more than 95$\%$ of the atoms in the atomic beam into the
metastable states. The laser is stabilized within a fraction of the
experimental linewidth. In the experiment we observed a linewidth of
the $^1$S$_0$-$^3$P$_1$ transition as low as 2~MHz. It is limited by
residual Doppler shifts in the divergent atomic beam, but it is
sufficient for good isotope selectivity.

\begin{figure}[tb]
\center
\includegraphics[width = 41 mm, angle = 270]{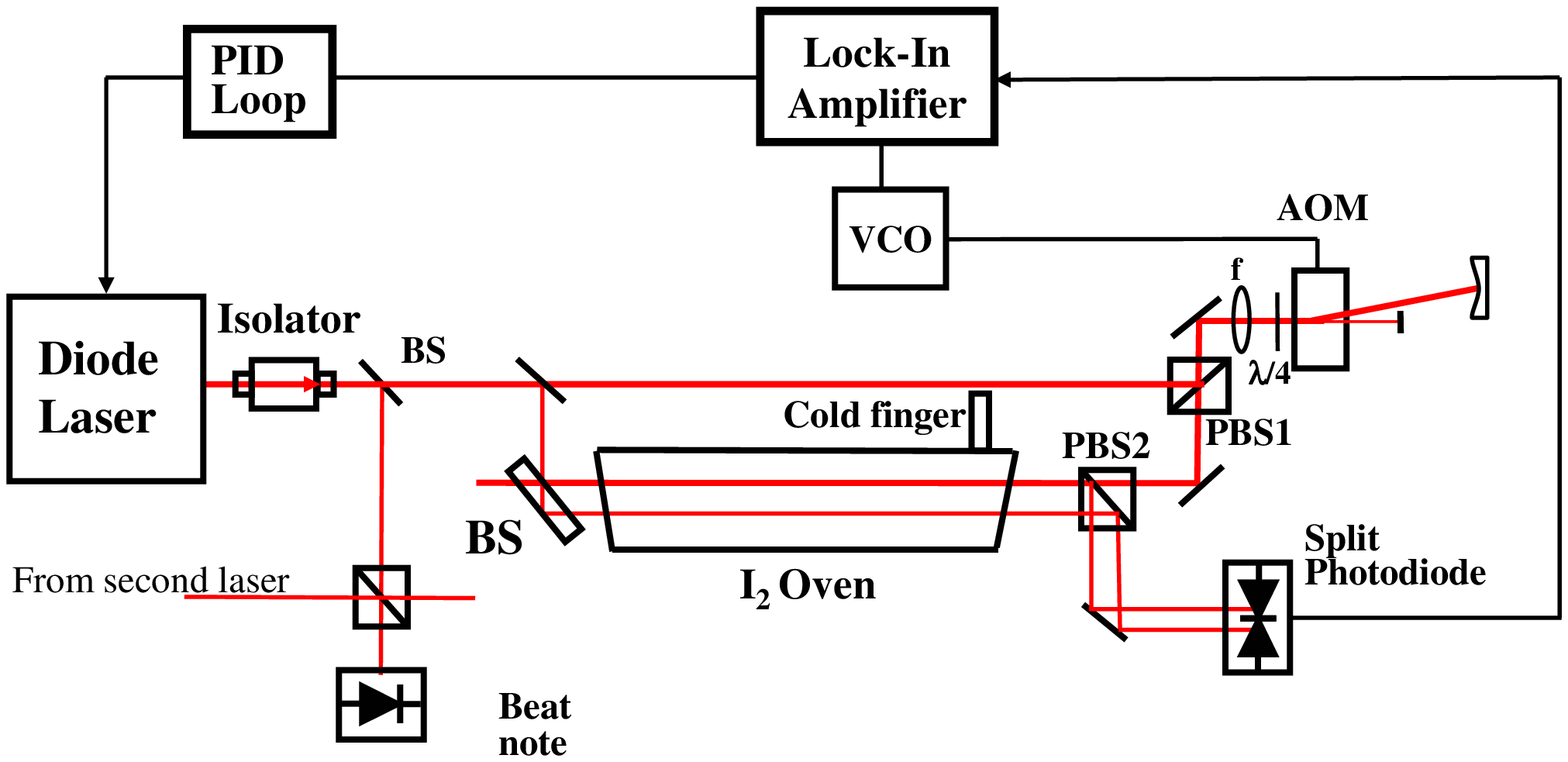}
\caption{Setup for frequency modulation saturated absorption
spectroscopy of iodine $^{127}$I$_{2}$. The laser beams have linear
polarization. The pump beam polarization is orthogonal to the
polarization of the two probe beams. The pump beam is frequency
modulated with a double passed AOM. The difference signal between
the two probe beams is detected on a balanced split photodetector.}
\label{iodinesetup}
\end{figure}

\begin{figure}[tb]
\center
\includegraphics[width = 85 mm]{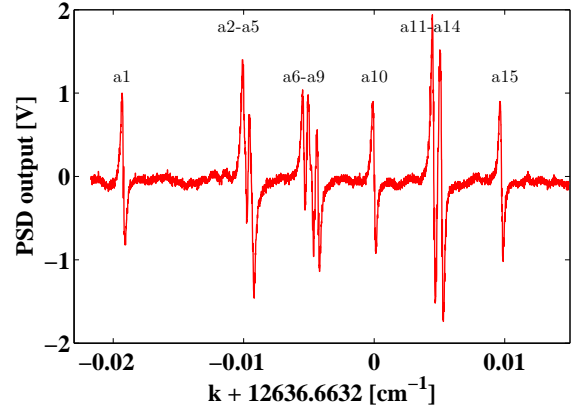}
\caption{Measured first-derivative hyperfine spectrum of P(52)(0-15)
transition in $^{127}$I$_{2}$ by frequency modulation saturated
absorption spectroscopy. We resolve the close lying multiplets from
which we determine a resolution of an individual line to 9~MHz. We
use the 'a1' component as the reference frequency.}
\label{iodinespec}
\end{figure}

In the course of the measurements we determined the absolute
frequency of the $^1$S$_0$-$^3$P$_1$ transition in $^{138}$Ba. The
transition is located -611(3)~MHz below the 'a1' iodine line, which
is at 12636.64357(1) cm$^{-1}$ \cite{GERSTENKORN1982,knockel2008}.
Our determination is a factor 100 better and within the uncertainty
of 0.001 cm$^{-1}$ of previous measurements \cite{curry2004}.

\subsection{Laser spectroscopy of D-states}

We use custom built fiber laser systems by KOHERAS  for probing the
$^3$D-states. The laser wavelengths can be temperature tuned by
about 0.4~nm and piezo tuned by about 0.3~nm. For changing the laser
frequency during the measurements we used the piezo, while we kept
the temperature constant. The observed thermal drift of the lasers
was less than 50 MHz/h. The transitions are always detected through
fluorescence at $\lambda_1$ from the $^{1}$P$_{1}$~-~$^{1}$S$_{0}$
transition which can be detected with a standard photomultiplier
tube (Hamamatsu R7205-01).

%The isotope shifts and resolution of the hyperfine structure of the
%odd isotopes of the $^{3}D_{1}$ and $^{3}D_{2}$ levels are carried
%out.

\section{Isotope shift of the $^3$D$-^1$P$_1$ transitions}

We investigate the isotope shifts of the transitions from the
$^3$D$_1$ and $^3$D$_2$-state, because they are sensitive to
electron correlations and to coupling to nuclear parameters. The
$^3$D$_1-^1$P$_1$ transition is the weakest among the three
repumping transitions for laser cooling of Ba. In the experimental
procedure, the known isotope shift \cite{GRUNDEVIK1983} and
hyperfine structure \cite{PUTLITZ1963} of the $^{3}$P$_{1}$ level
(Tab.~\ref{isoshiftpump}) were used to select different isotopes.
The measurements are summarized in Table \ref{3d1isotopeshift}.

\subsection{Even isotopes}

\begin{figure}[h]
\includegraphics[width = 100 mm, angle=270]{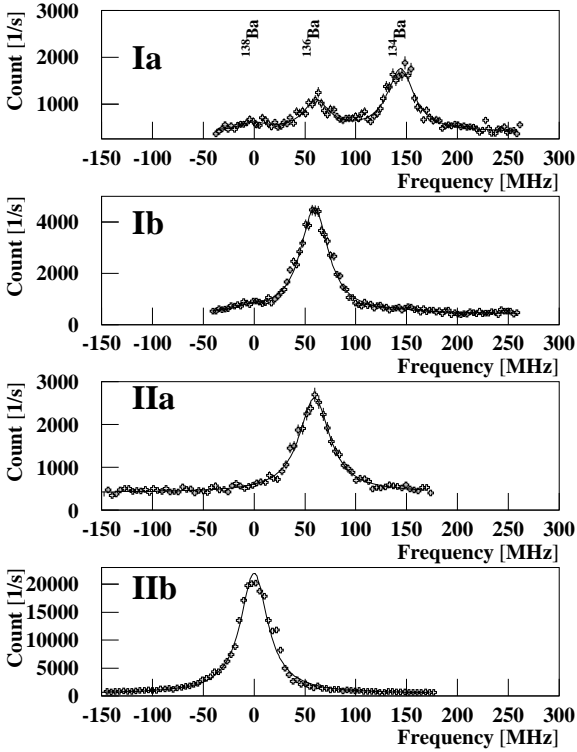}
\caption{Isotope shift for $^{138}$Ba, $^{136}$Ba, $^{134}$Ba at
wavelength $\lambda_{ir3}$. Spectra Ia and Ib, respectively IIa and
IIb, are taken simultaneously. (Ia) Selecting $^{134}$Ba. Because
the isotope shift at wavelength $\lambda_2$ of the pumping
transition is small, we populate also the other isotopes at the same
time. (Ib) and (IIa) Selecting $^{136}$Ba and (IIb) $^{138}$Ba. The
strength of the signal scales with the natural abundance of the
selected isotope. A Lorentzian lineshape is fitted to the
resonances. The frequency can be given relative to the transition in
$^{138}$Ba.} \label{fig_isoshifteven}
\end{figure}

Because of the absence of nuclear spin there is no hyperfine
structure for even isotopes. For measurements with these systems the
frequency of the diode laser at $\lambda_2$ is toggled between two
different isotopes by switching the frequency $\nu_{AOM}$. In this
way we probe both isotopes before we change the frequency of the
fiber laser at $\lambda_{ir3}$ (Fig.~\ref{fig_isoshifteven}). The
fitted linewidth is 33(3)~MHz for all resonances. The separation of
the resonances for the different isotopes at $\lambda_{ir3}$ is of
the order of the natural linewidth. Without isotope selective
population of the D-states the determination of the line center
would require much higher statistics. The signal amplitude is
proportional to the natural abundance of the isotopes. We find from
the peak intensities the abundance ratio $^{138}$Ba~:~$^{136}$Ba to
10.0(4):1, and $^{136}$Ba~:~$^{134}$Ba to 3.1(2):1, which is in good
agreement with the natural abundances.

\subsection{Odd isotopes}

In the odd isotopes we have hyperfine structure multiplets. The
hyperfine levels in the triplet D-states are populated by locking
the diode laser to the $^1$S$_0$($F$~=~3/2)~-~$^3$P$_1$($F$~=~3/2)
transition, which is shifted by -925.6(1.0)~MHz relative to the
transition in $^{138}$Ba. The selection rules allow to populate the
$F$~=~1/2, 3/2 and 5/2 states of the $^{3}$D$_{1}$ level for
$^{137}$Ba and $^{135}$Ba.

\begin{figure}[tb]
\center
\includegraphics[width = 90 mm]{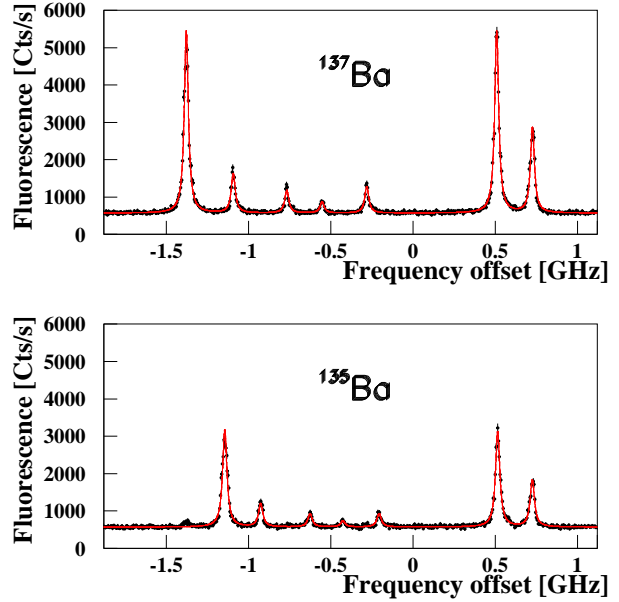}\\
\caption{Measured hyperfine spectra for the
$^{3}$D$_{1}$~-~$^{1}$P$_{1}$ transition for $^{137}$Ba
and$^{135}$Ba. The frequency axis is relative to the same transition
in $^{138}$Ba. Each frequency point was measured for 1~second.}
\label{3d1-1p1hfs}
\end{figure}

The hyperfine structure splitting of the $^{3}$P$_{1}$ state in Ba
isotopes with nuclear spin $I$~=~3/2 has been measured before with
high accuracy by optical-rf double-resonance, using a hollow cathode
lamp as a light source \cite{PUTLITZ1963}. The hyperfine structure
of the $^3$D$_1$ state was determined by an atomic beam magnetic
resonance method \cite{GUSTAVSSON1979}.

We have measured the hyperfine spectrum of $^{137}$Ba and $^{135}$Ba
isotopes for the $^{3}$D$_{1}$~-~$^{1}$P$_{1}$ transition
simultaneously by selecting the isotopes through optical pumping
using the intercombination line. The laser frequency at
$\lambda_{ir3}$ is scanned over the entire set of hyperfine lines
which are spreaded over more than 2~GHz (Fig.~\ref{3d1-1p1hfs}). All
the seven transitions can be fitted with a sum of Lorentzian
functions
\begin{equation}
F_\nu= C+\sum_{i=1}^{7}
\frac{A_i}{1+(\frac{\nu-\nu_i}{\Gamma/2})^2},
\end{equation}
where $C$ represents the background counts, $i$ is the index of the
hyperfine components, $A_i$ is the amplitude of each component,
$\Gamma$ is the full width at half maximum and $\nu_i$ is the center
of the hyperfine component. From the fit we obtain
$\Gamma$~=~20(1)~MHz, which is in good agreement with the natural
linewidth of the transition.

We derived the isotope shift between the isotopes $^{137}$Ba and
$^{135}$Ba from the measured hyperfine spectra. We determine the
difference frequencies $\Delta \nu_1$
(Tab.~\ref{137-135isotopeshift}) of all seven hyperfine components.
With the known hyperfine splittings of the $^{3}$D$_{1}$
\cite{GUSTAVSSON1979} and $^{1}$P$_{1}$-states
\cite{WIJNGAARDEN1995} we calculate the difference $\Delta \nu_2$
for all hyperfine transitions of the two isotopes. Thus, the isotope
shift $\Delta\nu_{IS}$ is
\begin{equation}
\Delta\nu_{IS} =  \Delta\nu_1 + \Delta \nu_2,
\end{equation}
where the difference in the hyperfine structure $\Delta\nu_2$ for
the two isotopes is
\begin{eqnarray}
\Delta \nu_2 & = & [\Delta\nu_{hfs}(^3D_1)
\Delta\nu_{hfs}(^1P_1)]_{137} \\
& & - [\Delta\nu_{hfs}(^3D_1) - (\Delta\nu_{hfs}(^1P_1)]_{135}.
\end{eqnarray}

\begin{table}[tb]
\center
\begin{tabular}{|c|r|r|c|}
   \hline
   \multicolumn{4}{|c|}{Odd isotopes: $^{3}D_{1}$-$^{1}P_{1}$
   transition}\\
   \hline\hline
   Transition & $\Delta \nu_1$ & $\Delta \nu_2$ & $\Delta\nu_{IS}$ \\
   & [MHz] & [MHz] & [MHz] \\
   \hline
1/2$\rightarrow$3/2 & -229.3(3.3)& 151.0(0.4) & -78.4(3.3)  \\
  1/2$\rightarrow$1/2 &-167.3(2.5) & 97.8(0.6) & -69.5(2.6) \\
  3/2$\rightarrow$5/2 & -143.2(2.6) &62.1(0.3)  & -81.1(2.6)  \\
 3/2$\rightarrow$3/2  & -128.3(4.2) &53.8(0.4)  & -74.4(4.2)  \\
  3/2$\rightarrow$1/2 & -76.2(2.2) & 0.7(0.6)& -75.5(2.3)  \\
 5/2$\rightarrow$5/2 & -6.1(0.6) & -68.5(0.3) & -74.6(0.7)  \\
 5/2$\rightarrow$3/2 & 1.2(1.0) & -76.7(0.4) &-75.5(1.1)  \\
   \hline
 \end{tabular}
\caption{Isotope shifts between $^{137}$Ba and $^{135}$Ba for the
6s5d~$^{3}$D$_{1}$~-~6s6p~$^{1}$P$_{1}$ transition. The measured
frequency difference $\Delta \nu_1$ of corresponding hyperfine
transitions and the difference in the hyperfine splitting $\Delta
\nu_2$ of the involved states from literature
\cite{WIJNGAARDEN1995,GUSTAVSSON1979} are given.}
\label{137-135isotopeshift}
\end{table}

We note that the error is smallest where the transitions nearly
coincide in frequency ((5/2$\rightarrow$5/2) and
(5/2$\rightarrow$3/2)). Here, the effect of the uncertainty of the
frequency calibration of the fiber laser is minimal. The average
over all seven values for $\Delta\nu_{IS}$ yields an isotope shift
of -75.3(5)~MHz between $^{137}$Ba and $^{135}$Ba.

\begin{table} \center
 \begin{tabular}{|c|c|c|}
   \hline
   Isotope pair & $^3$D$_1$-$^1$P$_1$ & $^3$D$_2$-$^1$P$_1$ \\
   & [MHz] & [MHz]  \\
   \hline
    138-136 & -59.3(6) & -63(1) \\
    138-134 & -144.1(1.0) & -143(1) \\
    136-134 & -84.8(8)& -80(1) \\
    138-137 & 114(4) & 69(3) \\
    138-135 & 39(4) &   \\
    137-135 & -75.3(5) & \\
    \hline
 \end{tabular}
 \caption{Isotope shifts for the
6s5d~$^{3}$D$_{1,2}$~-~6s6p~$^{1}$P$_{1}$ transition in Ba isotopes.
The uncertainty for the shift between odd and even isotopes is
systematically larger due to the uncertainty in the frequency
calibration of the fiber laser.} \label{3d1isotopeshift}
\end{table}

\section{Analysis of isotope shifts}

\begin{figure}[tbh]
\center
\includegraphics[width = 75 mm]{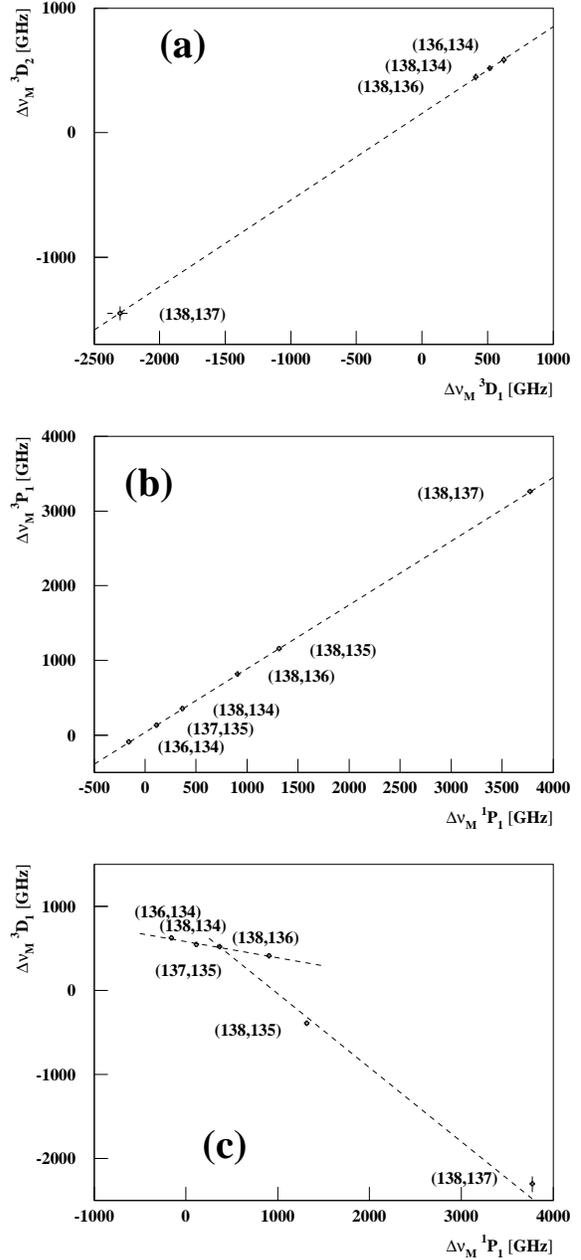}
\caption{King plot for the measured modified isotope shifts
$\Delta_{\rm M}$ of the $^3$D$_1$-$^1$P$_1$ and $^3$D$_2$-$^1$P$_1$
transitions in Ba isotopes. A linear correlation is found for the
transitions at the wavelengths  $\lambda_{ir2}$ and $\lambda_{ir3}$
(a) as well as at the wavelengths $\lambda_1$ and $\lambda_2$ (b).
For the the wavelength pair $\lambda_{ir2}$ and $\lambda_1$ we find
a deviation (c). There is a linear dependence for pairs of even-even
isotopes which differs from the relation for odd-odd and even-odd
pairs of isotopes. The data for $\lambda_1$ and $\lambda_2$ are
taken from \cite{GRUNDEVIK1983}.} \label{kingplot3d}
\end{figure}

The isotope shift of an atomic transition frequency between two
different isotopes of an atom with mass numbers $A_{1}$ and $A_{2}$
is traditionally written as a sum of normal mass shift
($\delta\nu_{NMS}$), specific mass shift ($\delta\nu_{SMS}$) and
field shift ($\delta\nu_{FS}$) \cite{WIJNGAARDEN1995}

\begin{equation}
\delta\nu_{IS}=\delta\nu_{NMS}+\delta\nu_{SMS}+ \delta\nu_{FS}.
\end{equation}

The normal mass shift arises from the reduced mass correction to the
energy levels in an atom and is

\begin{equation}
\delta\nu_{NMS} = \nu\frac{m_{e}}{m_{p}}
\frac{A_{1}-A_{2}}{A_{1}A_{2}},
\end{equation}

where $m_{e}$ is the electron mass, $m_{p}$ is the proton mass and
$\nu$ is the transition frequency. The specific mass shift is caused
by momentum correlation among orbital electrons and is

\begin{equation}
\delta\nu_{SMS} = F_{SMS}\, \nu\, \frac{A_{1}-A_{2}}{A_{1}A_{2}},
\end{equation}

where $F_{SMS}$ is a constant for a specific transition and it is
insensitive to the nuclear moments. It can be derived from theory.

The field shift arises due to the change in the spatial distribution
of the nuclear charge between the different isotopes and is

\begin{equation}
\delta\nu_{FS} = F\delta \langle r^{2} \rangle,
\end{equation}

where $F$ is a factor depending on the electronic configuration and
$\langle r^{2}\rangle$ is the expectation value of the square of the
radius of the charge distribution in the nucleus. This term provides
information on the overlap of the electron wavefunctions with the
nucleus.

Isotope shifts for different transitions are often compared using a
King plot~\cite{KING1963}. Here, the modified isotope shifts
$\Delta\nu_M$, which is given as

\begin{equation}
\Delta\nu_{M} = (\Delta\nu_{IS} -
\Delta\nu_{NMS})\times\frac{A_{1}A_{2}}{A_{1}-A_{2}},
\end{equation}

of a set of isotope pairs is plotted for two different transitions.
The King plot is expected to exhibit a linear dependence relating
the two isotope shift pair values. The slope gives the ratio of
field factors $F_{SMS}$ for the two transitions and the intercept is
related to the difference in specific mass shifts. Figure 7 gives
King plots for the isotope shifts of the $^3$D$_1$-$^1$P$_1$ and
$^3$D$_2$-$^1$P$_1$ transitions. A linear correlation  with a slope
of 0.696(20) is found for the modified isotope shift $\Delta
\nu_{\rm{M}}$ for the 5d-6p transitions at the wavelengths
$\lambda_{ir2}$  and $\lambda_{ir3}$ (Fig. 7a ). The linear
dependence for the 6s-6p transitions at wavelength $\lambda_1$ and
$\lambda_2$ has a slope of 0.852(6) (Fig. 7b ). It appears that in
the studied $^3$D$_{1,2}$-$^1$P$_1$ transitions the dominant single
electron contribution arises from a 5d-6p transition in contrast to
all other reported isotope shift measurements in this systems, where
stronger 5s-6p transitions were involved. The modified isotope shift
$\Delta \nu_{\rm M}$ for transitions with the same dominant single
electron contribution exhibit the expected linear correlation.

The situation is different for the 5d-6p transition at wavelength
$\lambda_{ir3}$ and the 6s-6p transition at wavelength $\lambda_1$.
For this transition pair we find a deviation from a linear
dependence (Fig. 7c). A linear correlation is only seen for the
subsets of even-even isotope pairs and subset of isotope pairs
involving odd isotopes. The slope for even-even pairs of isotopes is
-0.191(8), while the slope for odd-odd together with the even-odd
pairs is -0.88(2). This  indicates that the nuclear spin effects the
electron-electron correlations and gives rise to an additional
contribution to the isotope shift for atoms with nuclear spin.

We note that, similar to our observation, there are instances of
deviation from a linear function, e.g.,in Samarium isotopes [24].
The deviations may be due to electron-electron correlation in two
electron systems. The nuclear charge radius contribution is not
prominent. There are no theoretical calculations yet to compare our
results with.

\section{Conclusions}

In this work precision spectroscopy of the metastable D-states is
reported. It yielded the first time measurement of isotope shifts of
the $^3$D$_{1,2}$-$^1$P$_1$  transitions and optically resolved
hyperfine structure. The parameters obtained in the spectroscopy of
these states is essential input for effective laser cooling of heavy
alkaline earth systems. The established transition rates for these
weak transitions involved are sufficient to allow for repumping in a
laser cooling scheme.

The measurements contribute to the understanding of the atomic
structure of heavy alkaline earth elements. In particular they can
be exploited as input for atomic structure calculations which is in
progress in several groups \cite{privcomm}.

The isotope shifts of the 5d-6p $^3$D$_{1,2}$ -$^1$P$_1$ transitions
cannot be scaled from other measurements which were performed on
6s-6p transitions. In particular, the $^3$D-states seem to couple to
the nuclear parameters, which can be also seen in the large
hyperfine structure splitting, which is caused by so-called
core-polarization \cite{sahoo}. A solution of this discrepancy by
future theoretical work is needed in order to have the necessary
reliable wavefunction for heavy alkaline earth elements for, e.g.
enhancement factor calculations in edm search projects.

\section{Acknowledgements}

We would like to thank B. Sahoo for helpful discussions on the issue
of isotope shifts. This work was supported by the
\textit{Nederlandse Wetenschaptlijke Organisatie (NWO)} by and
NWO-VIDI grant, and the \textit{Stichting voor Fundamenteel
Onderzoek der Materie (FOM)} under programme 48 (TRI$\mu$P).

\end{document}